\documentclass[preprint,prc,aps,showpacs,showkeys,groupedaddress,floatfix]{revtex4}
\usepackage{epsfig}
\usepackage{dcolumn}
\usepackage{bm}
\usepackage[latin5]{inputenc}
\usepackage{graphics}
\usepackage{graphicx}
\usepackage{epsfig}
\usepackage{amssymb}
\usepackage{amsmath}
\begin{document}
\title{Arbitrary $\ell$-state solutions of the rotating Morse potential by the asymptotic iteration
method}
\author{O. Bayrak\dag \quad and I. Boztosun\ddag}
\affiliation{\dag Yozgat Faculty of Arts and Sciences, Department of
Physics, Erciyes University, Yozgat, Turkey, \\ \ddag Faculty of
Arts and Sciences, Department of Physics, Erciyes University,
Kayseri, Turkey}
\date{\today}
\begin{abstract}
For non-zero $\ell$ values, we present an analytical solution of the
radial Schr\"{o}dinger equation for the rotating Morse potential
using the Pekeris approximation within the framework of the
Asymptotic Iteration Method. The bound state energy eigenvalues and
corresponding wave functions are obtained for a number of diatomic
molecules and the results are compared with the findings of the
super-symmetry, the hypervirial perturbation, the Nikiforov-Uvarov,
the variational, the shifted 1/N and the modified shifted 1/N
expansion methods.
\end{abstract}

 \maketitle

\section{Introduction}
The Morse potential has raised a great deal of interest over the
years and has been one of the most useful models to describe the
interaction between two atoms in a diatomic molecule. It is known
that the radial Schr\"{o}dinger equation for this potential can be
solved exactly when the orbital angular quantum number $\ell$ is
equal to zero \cite{morse}. On the other hand, it is also known that
for $\ell\neq 0$, one has to use some approximations to find
analytical or semi-analytical solutions. Several schemes have been
presented for obtaining approximate solutions \cite{morales}. Among
these approximations, the most widely used and convenient one is the
Pekeris approximation \cite{Pekeris,fluge}, which is based on the
expansion of the centrifugal barrier in a series of exponentials
depending on the internuclear distance up to the second order. Other
approximations have also been developed to find better analytical
formulas for the rotating Morse potential. However, all these
approximations other than the Pekeris one require the calculation of
a state-dependent internuclear distance through the numerical
solutions of transcendental equations
\cite{depristo,elsum,morales2,bag}. In this respect, the rotating
Morse potential has so far been solved by the super-symmetry (SUSY)
\cite{susy,morales},  the Nikiforov-Uvarov method (NU)
\cite{Nifikorov,cuneyt}, the shifted and modified shifted 1/N
expansion methods \cite{bag,expansion} as well as the variational
method \cite{variational} using Pekeris approximations for $\ell\neq
0$. It is also solved by using the hypervirial perturbation method
(HV) \cite{killingbeck} with the full potential without Pekeris
approximation.

In this paper, our aim is to solve the rotating Morse potential
using a different and more practical method called, the asymptotic
iteration method (AIM) \cite{hakanaim1,hakanaim2} within the Pekeris
approximation and to obtain the energy eigenvalues and corresponding
eigenfunctions. In the next section, the asymptotic iteration method
(AIM) is introduced. Then, in section \ref{apply}, the
Schr\"{o}dinger equation is solved by the asymptotic iteration
method with the non-zero angular momentum quantum numbers for the
rotating Morse potential: The exact energy eigenvalues and
corresponding wave functions are calculated for the $H_{2}$, $HCl$,
$CO$ and $LiH$ diatomic molecules and AIM results are compared with
the findings of the SUSY \cite{morales}, the hypervirial
perturbation method (HV) \cite{killingbeck}, the Nikiforov-Uvarov
method (NU) \cite{cuneyt} and the shifted and modified shifted 1/N
expansion methods \cite{bag,expansion} as well as with the
variational method \cite{variational}. Finally, section
\ref{conclude} is devoted to the summary and conclusion.
\section{Basic Equations of the Asymptotic Iteration Method (AIM)}
We briefly outline the asymptotic iteration method here and the
details can be found in references \cite{hakanaim1,hakanaim2}. The
asymptotic iteration method is proposed to solve the second-order
differential equations of the form
\begin{equation}\label{diff}
  y''=\lambda_{0}(x)y'+s_{0}(x)y
\end{equation}
where $\lambda_{0}(x)\neq 0$ and s$_{0}$(x), $\lambda_{0}$(x) are in
C$_{\infty}$(a,b). The variables, s$_{0}$(x) and $\lambda_{0}$(x),
are sufficiently differentiable. The differential equation
(\ref{diff}) has a general solution \cite{hakanaim1}
\begin{equation}\label{generalsolution}
  y(x)=exp \left( - \int^{x} \alpha(x^{'}) dx^{'}\right ) \left [C_{2}+C_{1}
  \int^{x}exp  \left( \int^{x^{'}} (\lambda_{0}(x^{''})+2\alpha(x^{''})) dx^{''} \right ) dx^{'} \right
  ]
\end{equation}
if $k>0$, for sufficiently large $k$, we obtain the $\alpha(x)$
values from
\begin{equation}\label{quantization}
\frac{s_{k}(x)}{\lambda_{k}(x)}=\frac{s_{k-1}(x)}{\lambda_{k-1}(x)}=\alpha(x),
\quad k=1,2,3,\ldots
\end{equation}
where
\begin{eqnarray}\label{iter}
  \lambda_{k}(x) & = &
  \lambda_{k-1}'(x)+s_{k-1}(x)+\lambda_{0}(x)\lambda_{k-1}(x) \quad
  \nonumber \\
s_{k}(x) & = & s_{k-1}'(x)+s_{0}(x)\lambda_{k-1}(x), \quad \quad
\quad \quad k=1,2,3,\ldots
\end{eqnarray}
The energy eigenvalues are obtained from the quantization condition.
The quantization condition of the method together with equation
(\ref{iter}) can also be written as follows
\begin{equation}\label{kuantization}
  \delta_{k}(x)=\lambda_{k}(x)s_{k-1}(x)-\lambda_{k-1}(x)s_{k}(x)=0, \quad \quad
k=1,2,3,\ldots
\end{equation}

For a given potential such as the rotating Morse one, the radial
Schr\"{o}dinger equation is converted to the form of equation
(\ref{diff}). Then, s$_{0}(x)$ and $\lambda_{0}(x)$ are determined
and s$_{k}(x)$ and $\lambda_{k}(x)$ parameters are calculated. The
energy eigenvalues are determined by the quantization condition
given by equation (\ref{kuantization}). However, the wave functions
are determined by using the following wave function generator
\begin{equation}
y_{n}(x)=C_2 exp(-\int^{x}\frac{s_{k}(x^{\prime})}{\lambda
_{k}(x^{\prime})}dx^{\prime }) \label{generator1}
\end{equation}

\section{Calculation of the Energy Eigenvalues and Eigenfunctions}
\label{apply} The motion of a particle with the reduced mass $\mu$
is described by the following Schr\"{o}dinger equation:

\begin{equation}
\frac{-\hbar^{2}}{2\mu}\left(\frac{\partial^{2}}{\partial
r^{2}}+\frac{2}{r}\frac{\partial}{\partial r}+ \frac{1}{r^{2}}
\left[\frac{1}{\sin \theta}\frac{\partial}{\partial \theta} \left(
\sin \theta \frac{\partial}{\partial \theta} \right)
+\frac{1}{\sin^{2} \theta} \frac{\partial^{2}}{\partial \phi^{2}}
\right]+V(r) \right)\Psi_{n\ell m}(r,\theta,\phi)=E\Psi_{n\ell
m}(r,\theta,\phi)\label{sch}
\end{equation}
The terms in the square brackets with the overall minus sign are the
dimensionless angular momentum squared operator, ${\bf L}^{2}$.
Defining $\Psi_{n\ell m}(r,\theta,\phi)=u_{n\ell}(r)Y_{\ell
m}(\theta,\phi)$, we obtain the radial part of the Schr\"{o}dinger
equation:
\begin{eqnarray}
\left(\frac{d^{2}}{d{r}^{2}}+\frac {2}{r}\frac{d}{dr}
\right)u_{n\ell}(r)-\frac{2\mu}{\hbar^{2}}\left[V(r)+\frac{\ell(\ell+1)\hbar^{2}}{2
\mu r^{2}} \right]u_{n\ell}(r)+\frac{2 \mu
E}{\hbar^{2}}u_{n\ell}(r)=0 \label{radialr}
\end{eqnarray}
It is sometimes convenient to define $u_{n\ell}(r)$ and the
effective potential as follows:
\begin{equation}
u_{n\ell}(r)=\frac{R_{n\ell}(r)}{r}, \quad
V_{eff}=V(r)+\frac{\ell(\ell+1)\hbar^{2}}{2 \mu r^{2}}
\end{equation}
Since
\begin{equation}
\left(\frac{d^{2}}{d{r}^{2}}+\frac {2}{r}\frac{d}{dr}
\right)\frac{R_{n\ell}(r)}{r}=\frac{1}{r}\frac{d^{2}}{d{r}^{2}}R_{n\ell}(r)
\end{equation}
The radial Schr\"{o}dinger equation given by equation
(\ref{radialr}) follows that
\begin{equation}
\frac{d^{2}R_{n\ell}(r)}{d{r}^{2}}+\frac {2\mu}{\hbar^{2}}
\left[E-V_{eff} \right]R_{n\ell}(r)=0 \label{factorschrodinger}
\end{equation}
Instead of solving the partial differential equation (\ref{sch}) in
three variables $r$, $\theta$ and $\phi$, we now solve a
differential equation involving only the variable $r$, but dependent
on the angular momentum parameter $\ell$, which makes the solution
of this equation difficult for $\ell\neq 0$ or sometimes impossible within a
given potential.

\begin{figure}[h]
\includegraphics{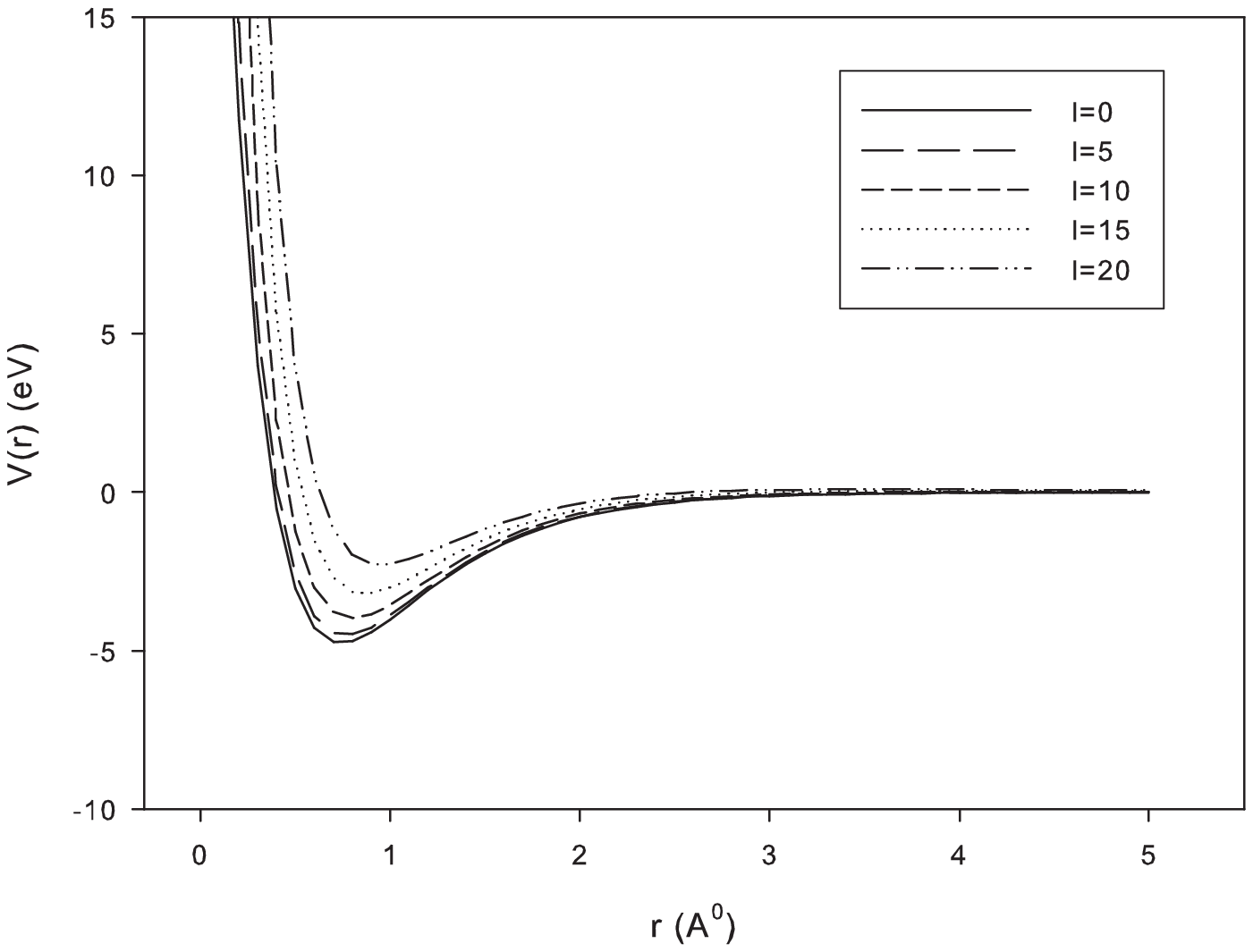}
\caption{The shape of the rotating Morse potential for $H_{2}$
diatomic molecule is plotted against the separation $r$ for
different orbital angular momentum quantum numbers.} \label{pot}
\end{figure}

The Morse potential we examine in this paper is defined as
\begin{equation}\label{morse}
  V_{Morse}(r)=D \left(e^{-2\alpha x}-2e^{-\alpha x} \right)
\end{equation}
with $x=(r-r_{e})/r_{e}$ and $\alpha=a r_{e}$. Here, $D$ and
$\alpha$ denote the dissociation energy and Morse parameter,
respectively. $r_{e}$ is the equilibrium distance (bound length)
between nuclei and $a$ is a parameter to control the width of the
potential well. For the $H_{2}$ diatomic molecule, the effective
potential, which is the sum of the centrifugal and Morse potentials,
is shown in figure \ref{pot} for various values of the orbital
angular momentum. The superposition of the attractive and repulsive
potentials results in the formation of a potential pocket, whose the
width and depth depend on the orbital angular momentum quantum
number for a given molecular potential. The potential pocket becomes
shallower as the orbital angular momentum quantum number $\ell$
increases, which also indicates that the number of states supported
by the potential decreases. This pocket is also very important for
the scattering case due to the interference of the barrier and
internal waves, which creates the oscillatory structure in the
cross-section. The effect of this pocket can be understood in terms
of the interference between the internal and barrier waves that
corresponds to a decomposition of the scattering amplitude into two
components, the inner and external waves~\cite{Lee78,Bri77,Bozosi}.

The effective potential together with the Morse potential for
$\ell\neq 0$ can be written as,
\begin{equation}\label{veff}
  V_{eff}(r)=V_{\ell}(r)+V_{Morse}(r)=\frac{\ell(\ell+1)\hbar^{2}}{2\mu r^{2}}+D \left(e^{-2\alpha x}-2e^{-\alpha x} \right)
\end{equation}
It is known that the Schr\"{o}dinger equation cannot be solved
exactly for this potential for $\ell\neq 0$ by using the standard
methods such as SUSY and NU. As it is seen  from
equation~(\ref{veff}), the effective potential is a combination of
the exponential and inverse square potentials, which cannot be
solved analytically. Therefore, an approximation has to be made: The
most widely used and convenient one is the Pekeris approximation.
This approximation is based on the expansion of the centrifugal
barrier in a series of exponentials depending on the internuclear
distance, keeping terms up to second order, so that the effective
$\ell$-dependent potential keeps the same form as the potential with
$\ell$=0 \cite{morales}. It should be pointed out, however, that
this approximation is valid only for low vibrational energy states.
In the Pekeris approximation, by change of the coordinates
$x=(r-r_{e})/r_{e}$, the centrifugal potential is expanded in a
series around $x=0$
\begin{equation}\label{expand}
V_{\ell}(x)=\frac{\ell(\ell+1)\hbar^{2}}{2\mu
r_{e}^{2}}\frac{1}{(1+x)^{2}}=\gamma(1-2x+3x^{2}-4x^{3}+...)
\end{equation}
where $\gamma=\frac{\ell(\ell+1)\hbar^{2}}{2\mu r_{e}^{2}}$. Taking
up to the second order degrees in this series and writing them in
terms of exponentials, we get

\begin{equation}\label{newexp}
\widetilde{V}_{\ell}(x)=\gamma\left(c_{0}+c_{1}e^{-\alpha
x}+c_{2}e^{-2 \alpha x} \right)
\end{equation}
In order to determine the constants $c_{0}$, $c_{1}$ and $c_{2}$, we
also expand this potential in a series of $x$
\begin{equation}\label{exp}
\widetilde{V}_{\ell}(x)=\gamma\left(c_{0}+c_{1}+c_{2}-(c_{1}+2c_2)\alpha
x+(\frac{c_{1}}{2}+2c_2)\alpha^2 x^2) \ldots \right)
\end{equation}
Comparing equal powers of equations (\ref{expand}) and (\ref{exp}),
we obtain the constants $c_{0}$, $c_{1}$ and $c_{2}$ as,
\begin{equation}\label{c}
    c_{0}=1-\frac{3}{\alpha}+\frac{3}{\alpha^{2}}, \quad  c_{1}=\frac{4}{\alpha}-\frac{6}{\alpha^{2}}, \quad
    c_{2}=-\frac{1}{\alpha}+\frac{3}{\alpha^{2}}
\end{equation}
Now, the effective potential with Pekeris approximation becomes,
\begin{equation}\label{veffpekeris}
\widetilde{V}_{eff}(x)=\gamma(c_{0}+c_{1} e^{-\alpha
x}+c_{2}e^{-2\alpha x})+D \left(e^{-2\alpha x}-2e^{-\alpha x}
\right)
\end{equation}
Instead of solving the radial Schr\"{o}dinger equation for the
effective potential given by equation (\ref{veff}), we solve the
radial Schr\"{o}dinger equation for the new effective potential
given by  equation (\ref{veffpekeris}) obtained by using the Pekeris
approximation. Inserting this effective potential equation
(\ref{veffpekeris}) into equation (\ref{factorschrodinger}) and
using the following \emph{ansatzs}
\begin{equation}\label{ansatz}
-\varepsilon^{2}=\frac{2\mu r_{e}^{2}}{\hbar^{2}}(E-\gamma c_{0}),
\quad \beta_{1}^{2}=\frac{2\mu r_{e}^{2}}{\hbar^{2}}(2D-\gamma
c_{1}), \quad \beta_{2}^{2}=\frac{2\mu r_{e}^{2}}{\hbar^{2}}(\gamma
c_{2}+D)
\end{equation}
The radial Schr\"{o}dinger equation takes the following form:
\begin{equation}\label{radyal}
 \frac{d^{2}R_{n\ell}(x)}{dx^{2}}+\left( -\varepsilon^{2}+\beta_{1}^{2} e^{-\alpha x}-\beta_{2}^{2} e^{-2\alpha x} \right) R_{n\ell}(x)=0
\end{equation}
If we rewrite equation (\ref{radyal}) by using a new variable of the
form $y=e^{-\alpha x}$, we obtain
\begin{equation}\label{trans}
\frac{d^{2}R_{n\ell}(y)}{dy^{2}}+\frac{1}{y}\frac{dR_{n\ell}(y)}{dy}+\left[-\frac{\varepsilon^{2}}{\alpha^{2}}\frac{1}{y^{2}}
+\frac{\beta_{1}^{2}}{\alpha^{2}}\frac{1}{y}-\frac{\beta_{2}^{2}}{\alpha^{2}}
\right]R_{n\ell}(y)=0
\end{equation}
In order to solve this equation with AIM for $\ell \ne 0$, we should
transform this equation to the form of equation (\ref{diff}).
Therefore, the reasonable physical wave function we propose is as
follows
\begin{equation}\label{wave}
R_{n\ell}(y)=y^{\frac{\varepsilon}{\alpha}}
e^{-\frac{\beta_{2}}{\alpha}y}f_{n\ell}(y)
\end{equation}
If we insert this wave function into the equation (\ref{trans}), we
have the second-order homogeneous linear differential equations in
the following form
\begin{equation}\label{aimschr}
\frac{d^{2}f_{n\ell}(y)}{dy^{2}}=\left(\frac{2\beta_{2}\alpha
y-2\varepsilon\alpha-\alpha^{2}}{y\alpha^{2}}\right)\frac{df_{n\ell}(y)}{d
y}+\left(\frac{2\varepsilon\beta_{2}+\alpha\beta_{2}-\beta_{1}^{2}}{y\alpha^{2}}
\right)f_{n\ell}(y)
\end{equation}
which is now amenable to an AIM solution. By comparing this equation
with equation (\ref{diff}), we can write the $\lambda_{0}(y)$ and
$s_{0}(y)$ values and by means of equation (\ref{iter}), we may
calculate $\lambda_k(y)$ and $s_k(y)$. This gives (the subscripts
are omitted):
\begin{eqnarray}\label{ini}
    \lambda_{0}&=&\left(\frac{2\beta_{2}y-2\varepsilon-\alpha}{\alpha y}\right) \nonumber \\
    s_{0}&=&\left(\frac{2\varepsilon\beta_{2}+\alpha\beta_{2}-\beta_{1}^{2}}{\alpha^{2}y}\right)\nonumber \\
    \lambda_{1}&=&{\frac {-3\,\beta_{2}\,\alpha\,y+6\,\alpha\,\varepsilon+2\,{\alpha}^{2}
-6\,y\varepsilon\,\beta_{2}-y{\beta_{1}}^{2}+4\,{\beta_{2}}^{2}{y}^ {2}+4\,{\varepsilon}^{2}}{{\alpha}^{2}{y}^{2}}}\nonumber \\
   s_{1}&=&2\,{\frac {\left (2\,\varepsilon\,\beta_{2}+\alpha\,\beta_{2}-{\beta_{1}}^{2}\right )\left (-\alpha+\beta_{2}\,y-\varepsilon\right )}{{
\alpha}^{3}{y}^{2}}} \\ \ldots \emph{etc} \nonumber
\end{eqnarray}
Combining these results with the quantization condition given by
equation (\ref{kuantization}) yields
\begin{eqnarray}
 \frac{s_0 }{\lambda _0 } & = & \frac{s_1 }{\lambda _1 }\,\,\,\,\,\, \Rightarrow
\,\,\,\,\,\,\varepsilon_{0}  =  -\frac{1}{2}\frac{\alpha\beta_{2}-\beta_{1}^{2}}{\beta_{2}} \nonumber \\
 \frac{s_1 }{\lambda _1 }  & = & \frac{s_2 }{\lambda _2 }\,\,\,\,\,\, \Rightarrow
\,\,\,\,\,\, \varepsilon_{1}=-\frac{1}{2}\frac{3\alpha\beta_{2}-\beta_{1}^{2}}{\beta_{2}} \nonumber \\
 \frac{s_2 }{\lambda _2 }  & = & \frac{s_3 }{\lambda _3 }\,\,\,\,\,\, \Rightarrow
\,\,\,\,\,\,\varepsilon_{2}=-\frac{1}{2}\frac{5\alpha\beta_{2}-\beta_{1}^{2}}{\beta_{2}} \\
\ldots \emph{etc} \nonumber
 \end{eqnarray}
When the above expressions are generalized, the eigenvalues turn out
as
\begin{equation}\label{energy}
\varepsilon_{n
\ell}=\frac{\beta_{1}^{2}-(2n+1)\alpha\beta_{2}}{2\beta_{2}},
\hspace{1cm} n=0,1,2,3,...
\end{equation}
Using equation (\ref{ansatz}), we obtain the energy eigenvalues
E$_{n\ell}$,
\begin{equation}\label{energyeigenvalues}
    E_{n\ell}=-\frac{\hbar^{2}}{2\mu r_{e}^{2}}\left[\frac{\beta_{1}^{2}}{2\beta_{2}}-(n+\frac{1}{2})\alpha\right]^{2}+\gamma c_{0}
\end{equation}
As it is seen that the energy eigenvalue equation is easily obtained
by using AIM. This is the advantage of the AIM that it gives the
eigenvalues directly by transforming the radial Schr\"{o}dinger
equation into a form of ${y}''$ =$ \lambda _0 (r){y}' + s_0 (r)y$.
In order to test the accuracy of equation (\ref{energyeigenvalues}),
we calculate the energy eigenvalues of the $H_{2}$, $HCl$, $CO$ and
$LiH$ diatomic molecules. The AIM results are compared with those
obtained by SUSY method \cite{morales} using original Pekeris
approximation, the hypervirial perturbation method (HV)
\cite{killingbeck}, the shifted 1/N and modified shifted 1/N
expansion methods \cite{bag} for the $H_{2}$ diatomic molecule in
Table \ref{Table1}. In Table \ref{Table2}, we show the same
comparison for the $HCl$ diatomic molecule. Furthermore, the AIM
results are compared with those obtained by NU method \cite{cuneyt},
shifted 1/N and modified shifted 1/N expansion methods \cite{bag}
for the $CO$ and $LiH$ diatomic molecules in Tables \ref{Table3} and
\ref{Table4}, respectively. As it can be seen from the results
presented in these tables that the AIM results are in good agreement
with the findings of the other methods.

After we find the energy eigenvalues, the following wave function
generator can be used to find $f_{n}(y)$ functions by using AIM
\begin{equation}
f_{n}(y)=exp(-\int^{y}\frac{s_{k}}{\lambda _{k}}dy^{\prime })
\label{generator}
\end{equation}%
where $n$ represents the radial quantum number and $k$ shows the
iteration number. Below, the first few $f(y)$ functions can be seen
\begin{small}
\begin{eqnarray}
f_{0}(y) &=&1 \\
f_{1}(y)&=&(2\alpha\beta_{2}-\beta_{1}^{2})\left(1-\frac{2\beta_{2} y}
{\alpha(\frac{\beta1^{2}-3\alpha\beta_{2}}{\alpha\beta_{2}}+1)}\right)\\
f_{2}(y)&=&(\beta_{1}^{2}-4\alpha\beta_{2})(\beta_{1}^{2}-3\alpha\beta_{2})
\left(1-\frac{4\beta_{2}y}{\alpha(\frac{\beta_{1}^{2}-5\alpha\beta_{2}}{\alpha\beta_{2}}+1)}
+\frac{4\beta_{2}^{2}y^{2}}{\alpha^{2}(\frac{\beta_{1}^{2}-5\alpha\beta_{2}}{\alpha\beta_{2}}+1)
(\frac{\beta_{1}^{2}-5\alpha\beta_{2}}{\alpha\beta_{2}}+2)}\right)\\
f_{3}(y)&=&(-4\alpha\beta_{2}+\beta_{1}^{2})(\beta_{1}^{2}-5\alpha\beta_{2})(\beta_{1}^{2}-6\alpha\beta_{2})
\left(1-\frac{6\beta_{2}y}{\alpha(\frac{\beta_{1}^{2}-7\alpha\beta_{2}}{\beta_{2}\alpha}+1)} \right. \nonumber \\
& + & \left. \frac{12\beta_{2}^{2}y^{2}}{\alpha^{2}(\frac{\beta_{1}^{2}-7\alpha\beta_{2}}{\alpha\beta_{2}}+1)
(\frac{\beta_{1}^{2}-7\alpha\beta_{2}}{\alpha\beta_{2}}+2)}-\frac{8\beta_{2}^{3}y^{3}}{\alpha^{3}
(\frac{\beta_{1}^{2}-7\alpha\beta_{2}}{\alpha\beta_{2}}+1)(\frac{\beta_{1}^{2}-7\alpha\beta_{2}}{\alpha\beta_{2}}+2)
(\frac{\beta_{1}^{2}-7\alpha\beta_{2}}{\alpha\beta_{2}}+3)}\right)\\
\ldots \emph{etc} \nonumber
\end{eqnarray}%
\end{small}
It can be understood from the results given above that we can write
the general formula for $f_{n}(y)$ as follows,
\begin{equation}
f_{n}(y)=(-1)^{n}\left(\prod\limits_{k=n}^{2n-1}(\beta_{1}^{2}-(k+1)\alpha\beta_{2})\right)
{_{1}}F_{1}
(-n,\frac{2\varepsilon_{n}}{\alpha}+1;\frac{2\beta_{2}y}{\alpha})
\end{equation}
Thus, we can write the total radial wave function as below,
\begin{equation}
R_{n\ell}=(-1)^{n}\left(\prod\limits_{k=n}^{2n-1}(\beta_{1}^{2}-(k+1)\alpha\beta_{2})\right)y^{\frac{\varepsilon_{n}}{\alpha}}e^{-\frac{\beta_{2}}{\alpha}y}
{_{1}}F_{1}
(-n,\frac{2\varepsilon_{n}}{\alpha}+1;\frac{2\beta_{2}y}{\alpha})
\end{equation}
When the hypergeometric function is written in terms of the Laguerre
polynomials, we get
\begin{equation}\label{wave1}
    R_{n\ell}=Ny^{\frac{\varepsilon_{n}}{\alpha}}e^{-\frac{\beta_{2}}{\alpha}y}
    L_{n}^{\frac{2\varepsilon_{n}}{\alpha}}\left(\frac{2\beta_{2}}{\alpha}y\right)
\end{equation}

 Where $N$ is the normalization constant and can be obtained from
$N^{2}\int\limits_{0}^{\infty}y^{\frac{2\varepsilon_{n}}{\alpha}}e^{-\frac{2\beta_{2}}{\alpha}y}
\left[L_{n}^{\frac{2\varepsilon_{n}}{\alpha}}(\frac{2\beta_{2}}{\alpha}y)\right]^{2}dy=1$
as below,
\begin{equation}\label{normalization}
    N=\frac{1}{n!}\left(\frac{2\beta_{2}}{\alpha}\right)^{\frac{\xi+1}{2}}\sqrt{\frac{(n-\xi)!}{n!}}
\end{equation}
where $\xi=\frac{2\varepsilon_{n}}{\alpha}$.

\section{Conclusion}
\label{conclude} We have shown an alternative method to obtain the
energy eigenvalues and corresponding eigenfunctions of the rotating
Morse potential using Pekeris approximation within the framework of
the asymptotic iteration method. The main results of this paper are
the energy eigenvalues and eigenfunctions, which are given by
equations (\ref{energyeigenvalues}) and (\ref{wave1}) respectively.
The energy eigenvalues are obtained for the $H_{2}$, $HCl$, $CO$ and
$LiH$ diatomic molecules. Our AIM results are compared with the
findings of the other methods such as the SUSY \cite{morales}, the
hypervirial perturbation method (HV) \cite{killingbeck}, the
Nikiforov-Uvarov method (NU) \cite{cuneyt} and the shifted and
modified shifted 1/N expansion methods \cite{bag,expansion} as well
as the variational method \cite{variational} in Tables \ref{Table1},
\ref{Table2}, \ref{Table3} and \ref{Table4}. The advantage of the
asymptotic iteration method is that it gives the eigenvalues
directly by transforming the radial Schr\"{o}dinger equation  into a
form of ${y}''$ =$ \lambda _0 (r){y}' + s_0 (r)y$. The wave
functions are easily constructed by iterating the values of $s_0(r)$
and $\lambda_0(r)$. The method presented in this study is a
systematic one and it is very efficient and practical. It is worth
extending this method to the solution of other interaction problems.

\section*{Acknowledgments} This paper is an output of the project supported by the
Scientific and Technical Research Council of Turkey
(T\"{U}B\.{I}TAK), under the project number TBAG-2398 and Erciyes
University (FBA-03-27, FBT-04-15, FBT-04-16). Authors would also
like to thank Dr. Ay\d{s}e Boztosun for the proofreading as well as
Dr. Hakan \c{C}ift\c{c}i for useful comments on the manuscript.

\newpage
\begin{table}
\begin{center}
\begin{tabular}{ccccccccccc}     \hline     \hline
&  & \mbox{AIM}& \mbox{SUSY}& \mbox{HV}&\mbox{Variational}&\mbox{Modified Shifted 1/N}&\mbox{Shifted 1/N} \\
\mbox{$n$}& \mbox{$\ell$}& \mbox{results}& \mbox{results} &
\mbox{results} &
\mbox{results}&\mbox{expansion results}&\mbox{expansion results} \\
\hline
 0     &     0   & -4.47601   &  -4.47601   & -4.47601    & -4.4758   &    -4.4760   &  -4.4749       \\
       &     5   & -4.25880   &  -4.25880   & -4.25901    & -4.2563   &    -4.2590   &  -4.2590       \\
       &     10  & -3.72193   &  -3.72193   & -3.72473    & -3.7187   &    -3.7247   &  -3.7247       \\
\\
 5     &     0   & -2.22052   &  -2.22051   &  -2.22051   &     -     &    -2.2205   &  -2.2038       \\
       &     5   & -2.04355   &  -2.04353   & -2.05285    &     -     &    -2.0530   &  -2.0525       \\
       &     10  & -1.60391   &  -1.60389   & -1.65265    &     -     &    -1.6535   &  -1.6526       \\
\\
 7     &     0   & -1.53744   &  -1.53743   & -1.53743    &     -     &    -1.5374   &  -1.5168       \\
       &     5   & -1.37656   &  -1.37654   & -1.39263    &     -     &    -1.3932   &  -1.3887       \\
       &     10  & -0.97581   &  -0.97578   & -1.05265    &     -     &    -1.0552   &  -1.0499       \\
\hline\hline
\end{tabular}
\end{center}
\caption{For the $H_{2}$ diatomic molecule, the comparison of the
energy eigenvalues (in eV) obtained by using AIM with other methods
for different values of $n$ and $\ell$. Potential parameters are
$D~=~4.7446eV$, $a=1.9425 (A^{0})^{-1}$, $r_{e}=0.7416A^{0}$,
${\hbar c}=1973.29 eV A^{0}$ and $\mu=0.50391$amu.} \label{Table1}
\end{table}

\begin{table}
\begin{center}
\begin{tabular}{ccccccccccc}     \hline     \hline
&  & \mbox{AIM}& \mbox{Variational}&\mbox{Modified Shifted 1/N}&\mbox{Shifted 1/N} \\
\mbox{$n$}& \mbox{$\ell$}& \mbox{results} &
\mbox{results}&\mbox{expansion results}&\mbox{expansion results} \\
\hline
0      &   0     &  -4.4356  &         -4.4360    &  -4.4355      &  -4.4352      \\
       &   5     &  -4.3968  &         -4.3971    &  -4.3968      &  -4.3967      \\
       &   10    &  -4.2941  &         -4.2940    &  -4.2940      &  -4.2939      \\
\\
5      &   0     &  -2.8051   &             -      &  -2.8046      &  -2.7727      \\
       &   5     &  -2.7721  &             -      &  -2.7718      &  -2.7508      \\
       &   10    &  -2.6847  &             -      &  -2.6850      &  -2.6712      \\
\\
7      &   0     & -2.2570   &             -      &  -2.2565      &  -2.2002      \\
       &   5     & -2.2263   &             -      &  -2.2262      &  -2.1874      \\
       &   10    & -2.1451  &             -      &  -2.1461      &  -2.1194      \\
\hline\hline
\end{tabular}
\end{center}
\caption{For the $HCl$ diatomic molecule, the comparison of the
energy eigenvalues (in eV) obtained by using AIM with other methods
for different values of $n$ and $\ell$. Potential parameters are
$D=37255cm^{-1}$, $a=1.8677(A^{0})^{-1}$, $r_{e}=1.2746A^{0}$,
${\hbar c}=1973.29 eV A^{0}$ and $\mu=0.9801045$amu.} \label{Table2}
\end{table}

\begin{table}
\begin{center}
\begin{tabular}{ccccccccccccc}     \hline     \hline
& & & &\mbox{AIM}& &\mbox{NU}& & \mbox{Variational}& &\mbox{Modified Shifted 1/N}& &\mbox{Shifted 1/N} \\
\mbox{$n$}& & \mbox{$\ell$}& &\mbox{results}& &\mbox{results}& &
\mbox{results}& &\mbox{expansion results}& &\mbox{expansion results}
\\ \hline
0       & & 0      & &-11.0915 & & -11.091         & &  -11.093       & &  -11.092   & &   -11.091    \\
       & & 5      & & -11.0844 & & -11.084         & &  -11.085       & &  -11.084   & &   -11.084    \\
       & & 10     & & -11.0653 & & -11.065         & &  -11.066       & &  -11.065   & &   -11.065    \\
\\
5       & & 0      & & -9.7952 & & -9.795          & &      -         & &   -9.795   & &    -9.788    \\
       & & 5      & & -9.7883  & & -9.788          & &      -         & &   -9.788   & &    -9.782    \\
       & & 10     & & -9.7701  & & -9.769          & &      -         & &   -9.770   & &    -9.765    \\
\\
7       & & 0      & &-9.2992  & & -9.299          & &      -         & &   -9.299   & &    -9.286    \\
       & & 5      & & -9.2925  & & -9.292          & &      -         & &   -9.292   & &    -9.281    \\
       & & 10     & & -9.2745  & & -9.274          & &      -         & &   -9.274   & &    -9.265    \\
\hline\hline
\end{tabular}
\end{center}
\caption{For the $CO$ diatomic molecule, the comparison of the
energy eigenvalues (in eV) obtained by using AIM with other methods
for different values of $n$ and $\ell$. Potential parameters are
$D=90540cm^{-1}$, $a=2.2994(A^{0})^{-1}$, $r_{e}=1.1283A^{0}$,
${\hbar c}=1973.29 eV A^{0}$ and $\mu=6.8606719$amu.} \label{Table3}
\end{table}

\begin{table}
\begin{center}
\begin{tabular}{ccccccccccccc}     \hline     \hline
& & & &\mbox{AIM}& &\mbox{NU}& & \mbox{Variational}& &\mbox{Modified Shifted 1/N}& &\mbox{Shifted 1/N} \\
\mbox{$n$}& & \mbox{$\ell$}& &\mbox{results}& &\mbox{results}& &
\mbox{results}& &\mbox{expansion results}& &\mbox{expansion results}
\\ \hline
0   & & 0      & & -2.4289 & & -2.4287         & &  -2.4291      & &  -2.4280      & &     -2.4278         \\
   & & 5      & &  -2.4013 & & -2.4012         & &  -2.4014      & &  -2.4000      & &     -2.3999         \\
   & & 10     & &  -2.3288 & & -2.3287         & &  -2.3287      & &  -2.3261      & &     -2.3261         \\
\\
5   & & 0      & & -1.6477 & & -1.6476         & &     -         & &  -1.6402      & &     -1.6242         \\
   & & 5      & &  -1.6238 & & -1.6236         & &     -         & &  -1.6160      & &     -1.6074         \\
   & & 10     & &  -1.5607 & & -1.5606         & &     -         & &  -1.5525      & &     -1.5479         \\
\\
7   & & 0      & & -1.3776 & & -1.3774         & &     -         & &  -1.3682      & &     -1.3424         \\
   & & 5      & &  -1.3550 & & -1.3549         & &     -         & &  -1.3456      & &     -1.3309         \\
   & & 10     & &  -1.2958 & & -1.2957         & &     -         & &  -1.2865      & &     -1.2781         \\
\hline\hline
\end{tabular}
\end{center}
\caption{For the $LiH$ diatomic molecule, the comparison of the
energy eigenvalues (in eV) obtained by using AIM with other methods
for different values of $n$ and $\ell$. Potential parameters are
$D=20287cm^{-1}$, $a=1.1280(A^{0})^{-1}$, $r_e=1.5956A^{0}$, ${\hbar
c}=1973.29 eV A^{0}$ and $\mu=0.8801221$amu.} \label{Table4}
\end{table}

\begin{thebibliography}{99}
\bibitem{morse} P. M. Morse, Phys. Rev. {\bf 34} (1929) 57.
\bibitem{morales} D. A. Morales, Chem. Phys. Letters \textbf{394} (2004) 68.
\bibitem{Pekeris} C. L. Pekeris, Phys. Rev. {\bf 45} (1934) 98.
\bibitem{fluge} S. Fl\"{u}gge, \texttt{Practical Quantum Mechanics} Vol. I, Springer, Berlin, (1994).
\bibitem{depristo} A.E. DePristo, J. Chem. Phys. \textbf{74} (1981) 5037.
\bibitem{elsum} J.R. Elsum and G. Gordon, J. Chem. Phys. \textbf{76} (1982) 5452.
\bibitem{morales2} D.A. Morales, Chem. Phys. Lett. \textbf{161} (1989) 253.
\bibitem{bag}   M. Bag \emph{et al.}, Phys. Rev. {\bf A46}, (1992) 9.
\bibitem{susy} F. Cooper, A. Khare and U. Sukhatme, Phys. Rep. {\bf 251} (1995) 267.
\bibitem{Nifikorov} A. F. Nikiforov and V. B. Uvarov, \texttt{Special Functions of Mathematical Physics}, Birkh\"{a}user, Basel, (1988).
\bibitem{cuneyt} C. Berkdemir and J. Han, Chem. Phys. Letters, \textbf{409} (2005) 203.
\bibitem{expansion} T. Imbo and U. Sukhatme, Phys. Rev. Lett. {\bf 54} (1985) 2184.
\bibitem{variational} E. D. Filho and R. M. Ricotta, Phys. Lett. {\bf A269} (2000) 269.
\bibitem{killingbeck} J.P. Killingbeck, A. Grosjean and G. Jolicard, J. Chem. Phys. \textbf{116} (2002) 447.
\bibitem{hakanaim1} H. Ciftci, R. L. Hall and N. Saad, J. Phys. A: Math. Gen. {\bf 36} (2003) 11807.
\bibitem{hakanaim2} H. Ciftci, R. L. Hall and N. Saad, Phys. Lett. \textbf{A340} (2005) 388.
\bibitem{Lee78} S.Y. Lee, Nucl. Phys. {\bf A311} (1978) 518.
\bibitem{Bri77} D.M. Brink and N. Takigawa, Nucl. Phys. {\bf A279} (1977) 159.
\bibitem{Bozosi} I. Boztosun, Phys. Rev. C {\bf 66} (2002) 024610.
\end{thebibliography}
\end{document}